\documentclass[article]{aa}

\usepackage{natbib}
\bibpunct{(}{)}{;}{a}{}{,} % to follow the A&A style
\usepackage{graphicx}
\usepackage{txfonts}

\usepackage{hyperref}
\usepackage{multirow}

\newcommand{\euclid}{\textit{Euclid}\xspace}

\begin{document}

   \title{Stellar classification from single-band imaging\\ using machine learning}

   \subtitle{}

   \author{T. Kuntzer
          \inst{1}
          \and 
          M. Tewes
          \inst{2}
          \and
          F. Courbin\inst{1}
          }

   \institute{Laboratoire d'astrophysique, Ecole Polytechnique F\'ed\'erale de Lausanne (EPFL), Observatoire de Sauverny, CH-1290 Versoix, Switzerland  
   \and 
   Argelander-Institut f\"ur Astronomie, Auf dem H\"ugel 71, D-53121 Bonn, Germany
            }

   \date{Received 07 April 2016; accepted 29 April 2016}

\abstract{
Information on the spectral types of stars is of great interest in view of the exploitation of space-based imaging surveys. 
In this article, we investigate the classification of stars into spectral types using only the shape of their diffraction pattern in a single broad-band image. We propose a supervised machine learning approach to this endeavour, based on principal component analysis (PCA)
 for dimensionality reduction, followed by artificial neural networks (ANNs) estimating the spectral type. 
Our analysis is performed with image simulations mimicking the Hubble Space Telescope (\textit{HST}) Advanced Camera for Surveys (\textit{ACS}) in the F606W and F814W bands, as well as the \euclid\ \textit{VIS} imager. 
We first demonstrate this classification in a simple context, assuming perfect knowledge of the point spread function (PSF) model and the possibility of accurately generating mock training data for the machine learning.
We then analyse its performance in a fully data-driven situation, in which the training would be performed with a limited subset of bright stars from a survey, and an unknown PSF with spatial variations across the detector.
We use simulations of main-sequence stars with flat distributions in spectral type and in signal-to-noise ratio, and classify these stars into 13 spectral subclasses, from O5 to M5. 
Under these conditions, the algorithm achieves a high success rate both for \euclid and \textit{HST} images, with typical errors of half a spectral class.
Although more detailed simulations would be needed to assess the performance of the algorithm on a specific survey, this shows that stellar classification from single-band images is well possible. 
  }


   \maketitle

   \section{Introduction}

Traditional methods to infer the spectral type of stars rely, as the name suggests, on the analysis of expensive spectra or multi-band photometry. Knowledge of spectral types and stellar parameters such as mass and age for large numbers of stars is of course of direct interest for stellar population studies and to study the formation history of our Galaxy \citep[e.g.][]{Smiljanic2014, Yang2015, Ness2015}.

More indirectly, stellar classification is also relevant for the future space telescopes \euclid\footnote{\url{http://www.euclid-ec.org/}} \citep{Euclid} and \textit{WFIRST} \citep{WFIRST}, as a reliable classification improves the quality of the reconstruction of the wavelength-dependent Point Spread Function (PSF) \citep[e.g.,][]{Cypriano2010} and as accurate knowledge of the PSF is mandatory to reach the scientific requirements for the weak gravitational lensing surveys \citep[for \euclid\ see e.g.][]{Cropper2013, Massey2013}. 
The \emph{VIS} imaging instrument of \euclid\ will feature a single broad filter. While this is  needed to reach the required number density of galaxies \citep{Euclid} to measure cosmic shear with sufficient precision, broad-band imaging also implies a number of complications in measuring galaxy shapes \citep{Voigt2012, Semboloni2013}. In addition, aside from the chromatic dependence of the PSF, a notable indirect effect arises from the spatially variable abundance of stars with companions \citep{Kuntzer2016}. 
Stellar data from \euclid\ can provide a wealth of information and contribute to a possible extension of the ESA \textit{Gaia} catalogue as \textit{Gaia} will provide stellar spectra for stars down to magnitude 17 \citep{Bruijne2015}.

In this paper, we present a novel technique to estimate the stellar spectral type of spatially unresolved sources, based solely on their image shape in a single wide band. This is important to carry out a first classification on the optical data of \euclid\ quickly and even for faint stars, beyond the reach of \textit{Gaia} or with no multi-band photometry available. 
Our technique will also be useful to classify stars in archival images of the Hubble Space Telescope (\textit{HST}). These images were taken in only one filter and therefore function as a general-purpose tool for stellar work.

The method exploits the subtle differences in diffraction limited images of point sources with contrasting spectra. A broad filter is generally advantageous for this approach, as it accentuates these differences between sources with varying spectral slopes. For this first approach, we perform the classification of sources into spectral types through a regression of a continuous scalar parameter, $C_s$, that roughly represents an effective temperature and covers adjacent bins of different spectral types. For each source,  estimates for $C_s$ are predicted by artificial neural networks \citep[ANN, see, e.g.,][]{Bishop1995}, using coefficients from a Principal Component Analysis \citep[PCA,][]{Pearson1901} of the source image as input. These neural networks perform a supervised machine learning, via training on stars with known spectral types.

All the images used in our exploratory work are simulations of stars along the main sequence, as observed either with \euclid or the \textit{HST}. This allows for a controlled proof of concept. But importantly, using these simulations, we also demonstrate the proposed technique in a purely data-driven application. For this, we mimic a situation in which a training set, with known true spectral types, is obtained by high resolution spectroscopy. 
To emulate an incomplete sampling of the training stars, we set aside some of the stellar spectra and spatial locations within the focal plane of the instrument during the training phase. 
We then analyse the performance of the method on stars with a lower signal-to-noise (S/N) cut, a greater variety of spectral types, and suffering from reddening by extinction. This complex test probes the interpolation behaviour of the classifier, and gives a first assessment of the reliability of results that could be expected on real data.

This article is organised as follows: 
we detail the algorithm and associated performance metrics in Section~\ref{sec:algos}. We then describe the preparation of the different simulated data sets for training and testing in Section~\ref{sec:data}.
In Section~\ref{sec:hyperopt}, we discuss the optimisation of the hyper-parameters. A proof-of-concept classifier and the performances of the classifiers for both \euclid\ and \textit{HST} are detailed in Section~\ref{sec:results}. Finally, section~\ref{sec:summary} summarises the work.

\section{Scheme and algorithms}  \label{sec:algos}

\begin{figure}[t]
\begin{center}
\includegraphics[width=1.\linewidth]{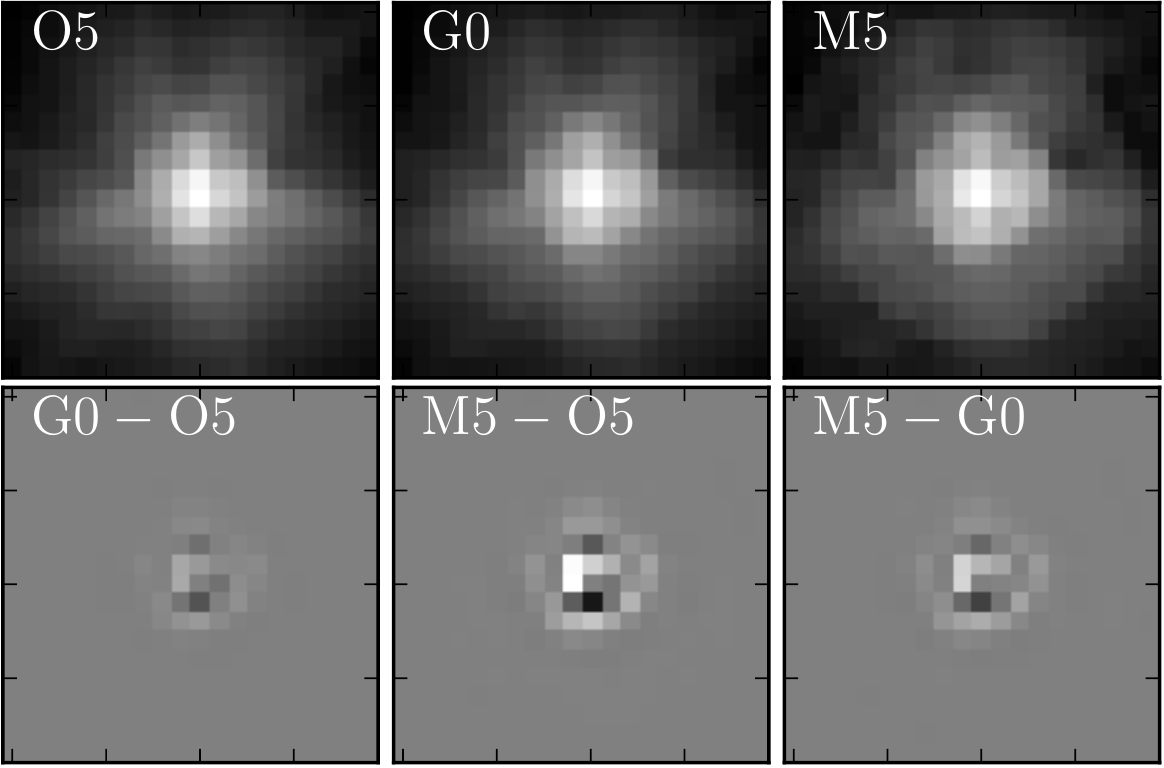}
\caption{\label{fig:different_stars} {\it Top:} simulated stellar images of different spectral types, as seen by the \euclid \emph{VIS} imager, shown with a logarithmic flux scale. {\it Bottom:} differences between pairs of these images, shown with a linear flux scale. White is positive and black is negative. Note that for demonstration purposes, this illustration is highly idealised: the above stellar images do not contain any noise, and the profiles are centred at exactly the same position with respect to the pixel grid.
}
\end{center}
\end{figure}

\begin{figure*}[t]
\begin{center}
\includegraphics[width=1.\linewidth]{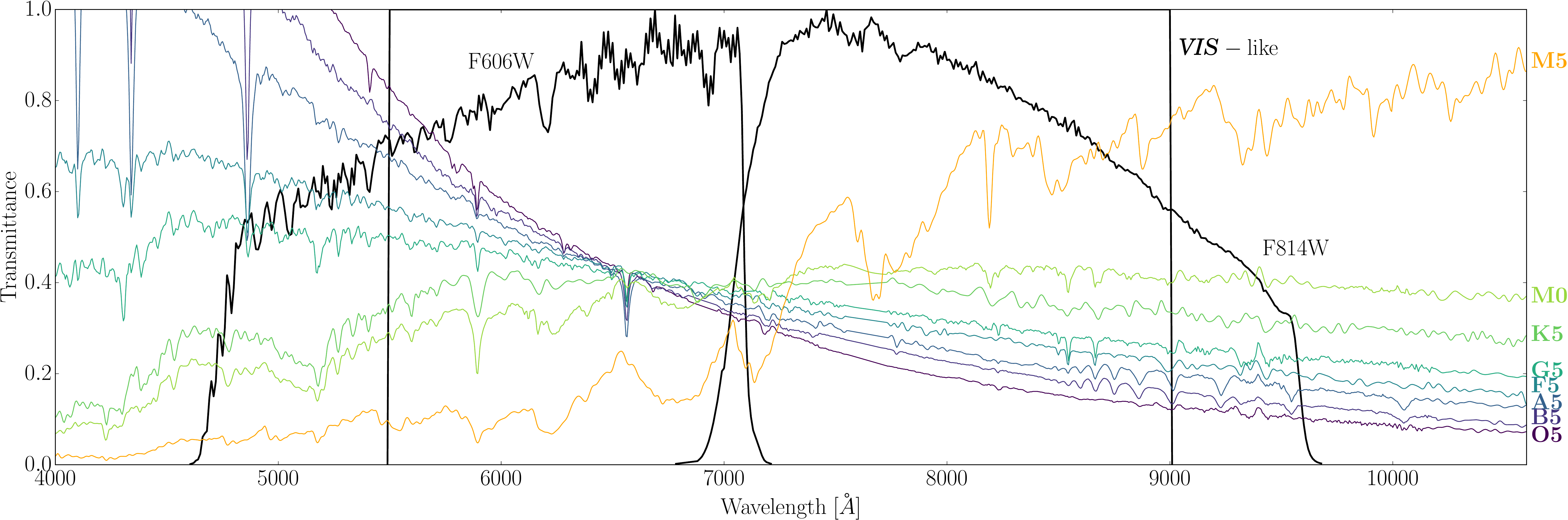}
\caption{\label{fig:spectra} Filter profiles of the three bands used in this work (\textit{VIS} from 550 to 900 nm, F606W and F814W) along with some stellar spectra of different types from \citet{Pickles1998}. For display purposes, the spectra have been normalised by their total flux in the \emph{VIS} band and plotted in arbitrary units of flux. Note that to simulate stellar images, we must integrate over the wavelength-dependent PSF models, using spectra in units of photon number counts
}
\end{center}
\end{figure*}

The proposed method, which we refer to as single-band classification, takes advantage of the fact that the diffraction-limited PSF of a telescope varies with wavelength. The precise shape of a stellar image, integrated over an observing filter, is therefore dependent on the transmission profile of the filter, as well as the stellar spectrum within this profile (for an illustration, see Figs.~\ref{fig:different_stars} and \ref{fig:spectra}). Our single-band classifier exploits these shape differences to predict the spectral class of a star. In the following, we succinctly lay down the different steps of the classifier, before describing them in more detail.

\begin{enumerate}

 \item Pre-processing of the data\\
To analyse images from a space-based survey, a catalogue is first created. This is performed through the detection of all stars, or, more generally, unresolved objects. Square stamps centered on the objects are prepared and normalised. Note that in this work, we simulate all the data, and directly produce stamps of pure stellar nature.

 \item Dimensionality reduction\\
Instead of using the normalised pixel values of a stamp as input to the machine learning, the image information is compressed, in order to reduce the dimensionality of the problem. To do so, the stamp images are projected onto a common basis, and only the most significant components are retained. In the vocabulary of machine learning, this reduces each stellar image to a chosen number of ``features''.

 \item Classification\\
The goal of this step is to create a robust mapping from the features to the spectral class of each object, using supervised machine learning. As commonly done in machine learning, we use an ensemble (``committee'') of classifiers and compare their outputs to (1) increase the confidence in the results, (2) estimate the uncertainty of the classification, and (3) detect unclassifiable objects.

\end{enumerate}

\subsection{Dimensionality reduction}
\label{dimred}

As the images of stars with different spectra do undeniably share common structures, they can be reconstructed, up to their noise, using a combination of components that are defined on a basis highlighting the differences between these images.
Finding this basis and retaining only a number of elements that represent the data well enough is the aim of dimensionality reduction. To this effect, we use the principal component analysis (PCA) technique. This algorithm projects the data onto the most meaningful basis \citep[see, e.g.,][]{Shlens2014} that represents the input data. A useful feature of the PCA decomposition is that it naturally provides a mean to compare the importance of each dimension. Since the projection is made along axes of decreasing importance for the reconstruction of the original data, all dimensions of order greater than a $n_\text{PCA}$ cut-off threshold can be dismissed. PCA has the advantage of being non-parametric, so that no hyper-parameters must be fine-tuned other than the number $n_\text{PCA}$ of components to be retained. PCA is widely used in astronomy, for example in PSF reconstruction \citep[e.g.][]{Jarvis2004, Gentile2013} and in weak lensing catalogue post-processing \citep[e.g.][]{Niemi2015}, to study properties of objects.

In practice, we simply use all available stellar images to construct the PCA basis onto which each star can be projected. In our analysis we compare results obtained by retaining from 12 to 27 PCA-coefficients for each star. We use the implementation of PCA provided by {\tt scikit-learn} \citep{scikit-learn}.

Note that as an alternative to PCA, we have tried to feed a moments-based width-measurement of the light profile as well as fluxes in different apertures as input features to the classification step. However these simple attempts turned out to be less successful than the PCA reduction. Other dimensionality reduction techniques, such as independent analysis component (ICA) or manifold mapping \citep[e.g. NMF,][]{Ivezic2014}, can also be applied to this problem, but they are not retained here as early attempts hinted at their similar or worse performance for the problem of classifying single-band stellar images.

\subsection{Classification: the machine learning} \label{sec:classification}

At this stage, through the dimensionality reduction, each stellar image can be seen as a point in an $n_\text{PCA}$-dimensional feature space. Classification methods such as k-nearest neighbour (k-NN) or support vector machine (SVM) rely on the clustering of the data into groups with the same labels, that is, the same spectral type \citep[see][for an overview]{Ivezic2014}.
Due to the image noise and the imperfect centering of stars with respect to the pixel grid, the different spectral types do not form clear disjoint clusters in PCA space, but exhibit a noisy but continuous evolution of the features. This will be illustrated, in the projection of two PCA components, in Fig.~\ref{fig:pca_decomposition}. The distribution of labelled data suggests a regression of a continuous scalar parameter, $C_s$, whose value evolves along the spectral classes. Eventually, the predicted class of a star is determined via a binning of $C_s$.

\subsubsection{Artificial neural networks}

We propose the use of simple artificial neural networks (ANN) to perform this regression from feature space to $C_s$.
Feed-forward ANNs of perceptrons \citep{Bishop1995} consist of several nodes, each taking an input vector $\pmb{x}$ and returning a scalar output $h(\pmb{x},\pmb{w}, b)$ via the equation
\begin{equation} \label{eq:ANN}
 h(\pmb{x},\pmb{w}, b) = h\left(\sum_{i=1}^N w_i x_i + b\right),
\end{equation}
where $\pmb{w}$ and $b$ are the weights and the bias, respectively. The monotonic and continuous function $h(x)$ is the so-called activation function. For our application, we use the sigmoid activation function $h(x) = 1/(1+e^{-x})$, except for the last node of the network, which uses the identity $h(x) = x$.
The nodes in the ANN are arranged into one or more layers. In each layer, nodes treat the input data through equation \ref{eq:ANN} with different values for the weights and the bias. In general, this input $\pmb{x}$ of each node consists of the outputs of the nodes in the previous layer. Nodes of the first layer take the vector of features as input, and the single node of the last layer returns the estimate for $C_s$. The capacity of a neural network to represent intricate dependencies depends on the number of nodes, and how these nodes are distributed into different layers. Choosing the number of nodes per layer and the number of layers is not straightforward, and we explore different combinations of number of layers and number of nodes per layer.
Layers that are not the input layer nor the output are called hidden layers.

For a given and fixed network structure, training of the ANN aims at finding optimal values of the weights and biases of each node, in order to minimise a cost function between the estimated and known true $C_s$ values of a training set where $C_s$ encodes the true spectral type (see section~\ref{sec:spectral_types}). We use the typical least-square cost function to evaluate the goodness of fit.

Various implementations of the multilayer perceptron could be used for the purpose of this study. We use the Fast Artificial Neural Network Library ({\tt FANN}) by \citet{Nissen2003}. We have also tried the {\tt SkyNet} implementation \citep{Graff2014}, yielding very similar results. As we do not aim to compare implementations of ANNs in the scope of this paper, we only report results obtained with {\tt FANN} in the following sections.
Other algorithms such as random forests (RF) can be applied here. Simple tests carried out with RF instead of ANNs yielded similar performance.

\subsubsection{Committees for better robustness and anomaly detection}

Due to the complexity of an ANN training, and random initialisation of weights and biases, the final values of the parameters obtained through the minimisation of the cost function are not deterministic. A training attempt can also remain trapped in a poor local minimum of the cost function. 

To address these difficulties, and increase the prediction accuracy, several independent ANNs, forming a so-called committee, can be trained individually \citep{Bishop1995}. This allows us to reject the worst training failures, based on the cost function performance achieved on the training set, and retain only the $n_c$ best committee members. When analysing unknown data, the different predictions from these retained committee members can be averaged, to yield a robust combined estimate for each object. A large variance of predictions is an indication that the unknown object was not represented in the training data. Another possible response to such an anormal object would be an ensemble of predictions that fall far from the range of known values of $C_S$. The committee approach increases the confidence in detecting anomalies \citep{Nguyen2015}. In the present context, such anomalies could range from slightly resolved objects such as small galaxies, to unresolved objects with unusual spectra (binary stars, quasars) or too-noisy data. In the following section, we define how exactly these outliers are identified.

\subsubsection{Classification into spectral types and anomalies} 
\label{sec:spectral_types}

In this paper we consider the classification into a set of 13 separate classes of stellar spectra, with a discretisation of ``half'' a spectral type: $\{$O5, B0, B5, A0, A5, F0, F5, G0, G5, K0, K5, M0, M5$\}$. We define the continuous parameter $C_s$ by attributing a sequence of numerical values to these classes, in steps of $0.5$. Training stars of type 05 get a true $C_s$ of 1.5, and $C_s(\textrm{B0}) = 2.0, C_s(\textrm{B5}) = 2.5, \ldots, C_s(\textrm{M5}) = 7.5$.

For each unknown object to be analysed, the combined average $C_s$-estimates from the retained well-trained committee members determines the classification: O5 if $1.25 < \langle C_s \rangle \le 1.75$, B0 if $1.75 < \langle C_s \rangle \le 2.25$, and so on until M5 with $7.25 < \langle C_s \rangle \le 7.75$. We refer to an estimation error of 0.5 on the $C_s$-scale as an error of half a spectral type.

In addition, if the variance of the individual $C_s$-estimates is larger than 1.0 or if $\langle C_s\rangle$ is out of range, we classify the object as an anomaly.

\begin{table*}[t!]
\caption{\label{tab:sets} Summary of the characteristics of the three data set families.}
\centering
\begin{tabular}{lccccccccl}
\hline
\hline
  Data set & \# Spectra & \# PSF & $A_v$ & S/N$_\text{Euclid}$ & S/N$_\text{HST F606W}$ & S/N$_\text{HST F814W}$ & \# Stars \\\hline
  Training & 13 & 10 & 0  & 50-400 & 120-1000 & 200-1000 & $\sim32000$\\
  Validation & 13 & 10 & 0 & 50-400 & 120-1000 & 200-1000 & $\sim20000$\\
  Test & 27 & 600 & 0.3 & 20-400 & 80-400 & 150-400 & $\sim20000$\\
\hline
\end{tabular}
\tablefoot{For each data set we give the number of different spectral templates, the number of different spatial positions on the detector, the maximum extinction $A_v$ (in magnitude), the considered S/N ranges, and the number of simulated stars. The extinction in our simulated data is randomly drawn between $A_{v,\text{min}}=0$ and $A_v$ (see text).}
\end{table*}

\subsection{Metrics to quantify the classification performance}\label{metrics}
To analyse the performance of the single-band classifier applied to a large sample of objects, we introduce a set of simple metrics. We describe them below.

\begin{itemize}

\item The \emph{confusion matrix}, whose elements $M_{ij}$ correspond to the relative abundance of the estimated spectral type $i$ given the true spectral type $j$. Correctly classified objects contribute to the diagonal terms of the matrix, while classification errors are represented by the off-diagonal elements. The distribution of the objects in the confusion matrix can reveal systematic biases and give a detailed overview of the classification errors.

\item The $F_1$\emph{-score} is a metric which summarises further the performance to one scalar value. For a binary classification, the $F_1$-score is defined by 
\begin{equation}
F_1 = \frac{2\mathrm{TP}}{2\mathrm{TP}+\mathrm{FN}+\mathrm{FP}},
\end{equation}
where $\mathrm{TP}$, $\mathrm{FN}$, and $\mathrm{FP}$ are the numbers of true positive, false negative, and false positive classifications, respectively.
We compute $F_1$ individually for each of the spectral types, and average these results to get a single $F_1$-score describing the overall classification performance. An error-free classification corresponds to $F_1 = 1$, and imperfect classifications reach lower scores. Note that this is a very strict measure of performance, as it will consider an object to be wrongly classified  if the estimate falls into a class immediately adjacent to the true spectral type. In other words, given the spectral classes used in this work, it even penalises errors corresponding to only half a spectral type (e.g., G5 instead of G0).

\item The success rate $S$ is the classification accuracy including a tolerance of one class (i.e., half a spectral type). In practice, $S$ is the trace of the confusion matrix plus the sum of the elements directly above and below the main diagonal, divided by the overall number of classified objects. In this paper, we optimise the configuration of the single-band classifier according to this success rate $S$.

\end{itemize}

\section{Simulated data} \label{sec:data}

In this section, we describe the preparation of synthetic data sets mimicking stellar images obtained by the \textit{HST} and \euclid. We first present the structure and methodology that we use for creating the mock images, and then discuss the telescope-specific tools to produce realistic images.

\subsection{Training, validation, and testing}
In line with machine learning practices \citep[e.g.][]{Hastie2009}, for each observational setup to be simulated, we generate a group of three disjoint data sets, all with known true spectral type. A similar structure could be adopted to split the subset of data with known spectral classification when working with real observations.

\begin{itemize}

\item First, a training set is needed, on which the neural networks learn by adjusting their weights and biases. Potentially, over-fitting of the neural network parameters could lead to exceedingly high apparent performances on this training set. Over-fitting arises when the dimensionality reduction or/and the neural networks become too specific to the data, for example, by fitting the noise contained in the training set.

\item The validation set is not seen by the neural networks during the optimisation of their parameters. By comparing the classification performance on the training set and the validation set, over-fitting of the neural networks can be detected.
If no over-fitting is detected, and if this validation set is large enough, it can in turn be used to optimise the hyper-parameters of the machine learning algorithm, such as, in the case of this work, the number $n_\text{pca}$ of PCA coefficients and the size of neural networks.
 
\item Finally, a test set is prepared, to independently test the performance of the optimised algorithm.

\end{itemize}

In the context of this paper, for some analyses we add additional astrophysical and observational complexity to the test set. Compared to the training and validation sets, we include fainter stars, more variants of the PSF corresponding to different spatial positions on the detector, additional stellar spectra, and wavelength-dependent extinction by dust. Thereby, our test sets can also be used to explore the performance of the classifier on significantly more complex data, mimicking a purely data-driven approach in which the training could not be performed on fully representative samples.

\begin{table*}[t!]
\caption{\label{tab:optimisation} Optimal configurations of the single-band classifiers, yielding best results on the validation sets.}
\centering
\begin{tabular}{l|cccc|cc|cc}
\hline
\hline
  Observational setup &  $n_\text{PCA}$ & $n_\text{hn}$ & $n_l$ & $n_\text{c}$ & $F_{1,\text{va}}$ & $F_{1,\text{tt}}$ & $S_\text{va}$ & $S_\text{tt}$ \\\hline
  \euclid & 24 & 26 & 2 & 48 & 0.75  & 0.42 & 0.98 & 0.90 \\
  {\it HST} F606W & 27 & 29 & 3 & 24 & 0.57 &  0.30 & 0.94 & 0.68  \\
\hline
\end{tabular}
\tablefoot{
The hyper-parameter $n_\text{PCA}$ is the number of retained PCA components, $n_\text{hn}$ is the number nodes in the hidden layers of the ANN, $n_l$ is the number of hidden layers, and $n_c$ is the number of ANNs retained in the committee (out of the 96 trained). The $F_1$ score and the success rate $S$ are given for the validation (va) and test (tt) sets. If the output catalogues were randomly drawn, the metrics would be $F_1 \approx 0.07$ and $S\approx 0.21$.}
\end{table*}

\begin{figure*}
 \centering
 \includegraphics[width=.9\linewidth]{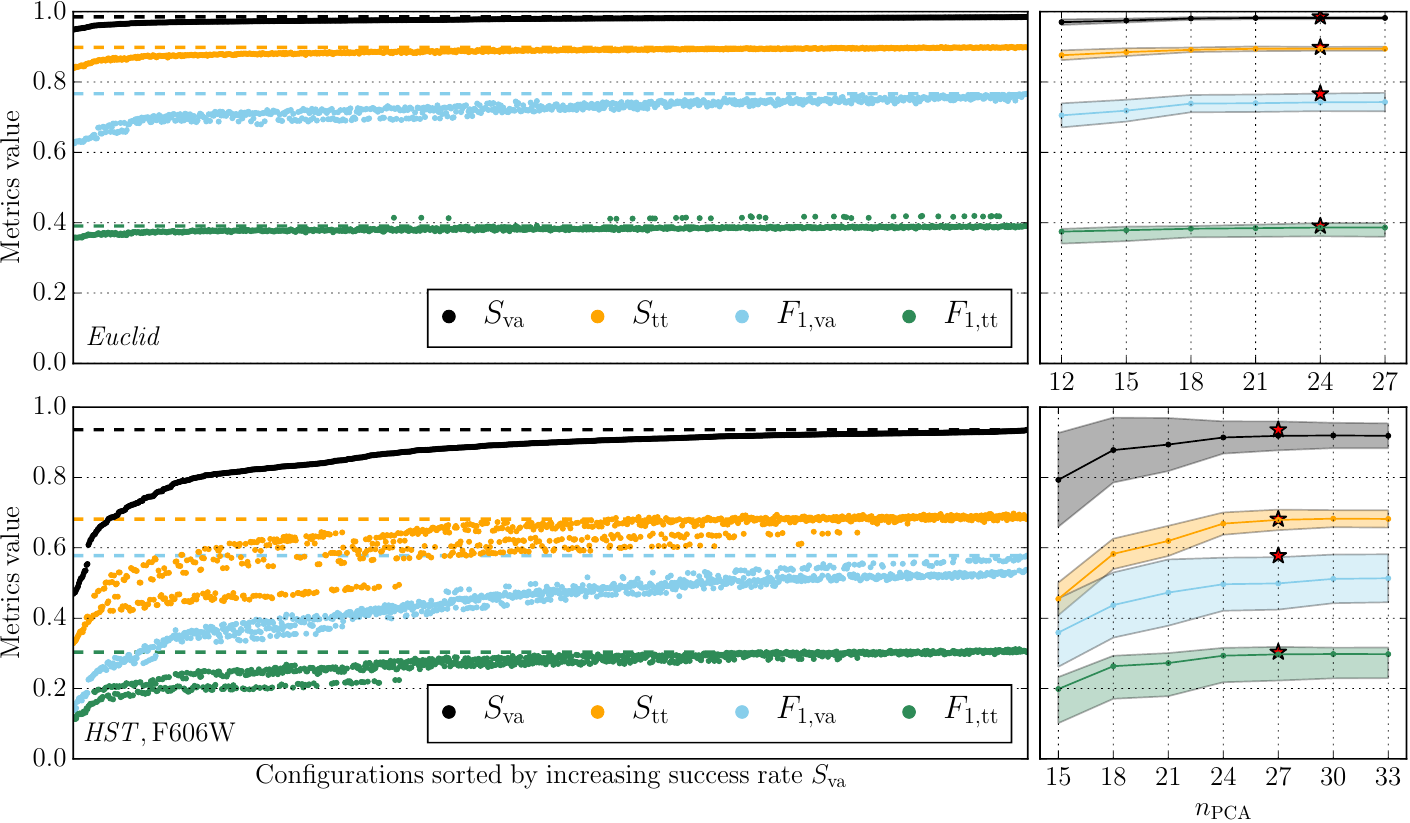}
  \caption{\label{fig:configs} Classification performances achieved by the different hyper-parameter combinations. The top panels are for the \euclid\ data sets, while the bottom panels are for the \textit{HST} F606W filter. The left-hand plots show the performance of the all tested configurations in terms of the $F_1$-score and of the success rate $S$, for the validation (va) and test (tt) sets. The dashed lines show the performance of the best configuration in both cases, selected by the highest $S_\text{va}$ score. The right-hand plots depict the median metrics values for configurations with a given number of PCA components. The shaded regions depict the 1$\sigma$ envelope on the median. The red stars correspond to the optimal classifiers.}
\end{figure*}

\subsection{Mock stellar images: generalities}

We restrict the range of stellar spectra to main sequence spectra using the templates prepared by \citet{Pickles1998}.
A few of these  are shown in Fig.~\ref{fig:spectra}.
For all our data sets, we adopt flat uniform distributions of these spectral types and of the S/N. Inevitably, the global performance of the single-band classifier depends on the stellar distribution, as the different stellar type yield different performances. For real data, the stellar distribution would depend on the galactic coordinates \citep[]{Chabrier2003,Robin2004}.
Our choice of a flat distribution has the advantage that a sufficient number of stars can be drawn in each stellar type bin while maintaining a tractable total size of the data sets. Tests on flat distributions could be later weighted to predict results for arbitrary stellar distributions. The same arguments motivate our choice of working with flat S/N distributions.

Table \ref{tab:sets} summarises the characteristics of the three data sets, which we generate for each considered band and telescope.
For the training and the validation sets, we restrict the diversity of PSFs to 10 different spatial locations on the detector, and use only the 13 different spectra (two per spectral class with an exception as we start from O5) defining the classification. For the more complex test sets, we finely sample all detector positions, and use all spectral templates from the library provided by \citet{Pickles1998} (roughly four per spectral class). For the purpose of evaluating the performance metrics, the true spectral types of these templates are rounded to the nearest classification bin (e.g., M4 becomes M5). Finally, we also add the effect of reddening by dust to the test sets only, using a Milky Way extinction curve with $R_V=3.1$ and the extinction $A_v$ randomly chosen between 0 and 0.3, to reflect the typical visual extinctions for the sky of the \euclid\ weak lensing survey \citep{Cardelli1989, Schlegel1998, Schlafly2011}.

Instrument-specific codes then produce the image of the objects, according to the different bands, spectra, and fluxes. In all our simulations, objects are randomly mis-centered by up to half a pixel in each direction both on the $x$ and $y$ axes, to obtain a uniform coverage of the sub-pixel positions and simulate a non-interpolating stamp extraction from survey data.

\subsection{Simulated \euclid\ images}

The PSFs we use for the \euclid\ telescope \citep{Euclid} are simulated using the pipeline for the VIS instrument (P. Hudelot, private comm.) and consists of 600 PSFs at random spatial positions within the four central CCD chips of the VIS camera. Depending on the position on the detector, the measured axis ratio evolves from 1 to 1.15. Each PSF is a FITS datacube containing 100 wavelength slices, hence allowing us to accurately describe realistic SEDs. To produce stellar images for VIS we consider a top-hat window function between 550 and 900~nm. The pixel size is that of the VIS detector (no sub-sampling), that is $\Delta x = 0.1$\arcsec. 

The S/N range for the \euclid\ training and validation sets spans $50 < \text{S/N} < 400$, while the test set images have a lower S/N cut of $\text{S/N}=20$. The limiting AB magnitude for \euclid is $\text{V}\approx 24.5$, which corresponds to S/N\ $\approx 10$ \citep{Euclid}.

The number of training stars is of the order of 32000. The validation and test sets contain about 20~000 images. We observe that each of these samples is large enough to exclude any over-fitting when using machine learning methods.

\subsection{Simulating Hubble Space Telescope images}

For the \textit{HST} simulations, we simulate stellar images in the F606W and F814W bands of the Advanced Camera for Surveys \citep[\textit{ACS},][]{Ford1996, Sirianni2005}. Both bands have similar widths, but are centred on different wavelengths (see Figure \ref{fig:spectra}). The \textit{HST} bandwidths are both about 1.5 times smaller than the \euclid VIS band. In addition, using the actual throughput curves, instead of an idealised top-hat function, will also reduce the potential performance of the single-band classification. 
The images are produced via the {\tt TinyTim} software \citep{Krist2011} in its \textit{ACS} configuration (both CCDs are used), using the same template spectra from \citet{Pickles1998} as for the \euclid\ simulations.
The lower bound for the S/N range, S/N\ $=80$, corresponds to a limiting AB magnitude of Johnson $V\approx23.5$ for O5V stars and $V\approx24.1$ for M5V stars with an exposure time of one hour \citep{Avila2016}. For S/N\ $=1000$, (the higher bound of the training set), the corresponding limiting magnitudes are $V\approx19.5$ and $V\approx20$ for O5V and M5V stars respectively.

Our aim in simulating these two F606W and F814W bands is not to compare their performance as input to a single-band classifier. Any such comparison would only be possible given a particular scientific question, and for a particular stellar population. Instead, we adjust here the arbitrary S/N ranges so that our classifiers yield results of roughly similar quality from both bands. This demonstrates that the single-band classification is possible both with F606W and F814W images.

\begin{figure}[t!]
\centering
\includegraphics[width=.9\linewidth]{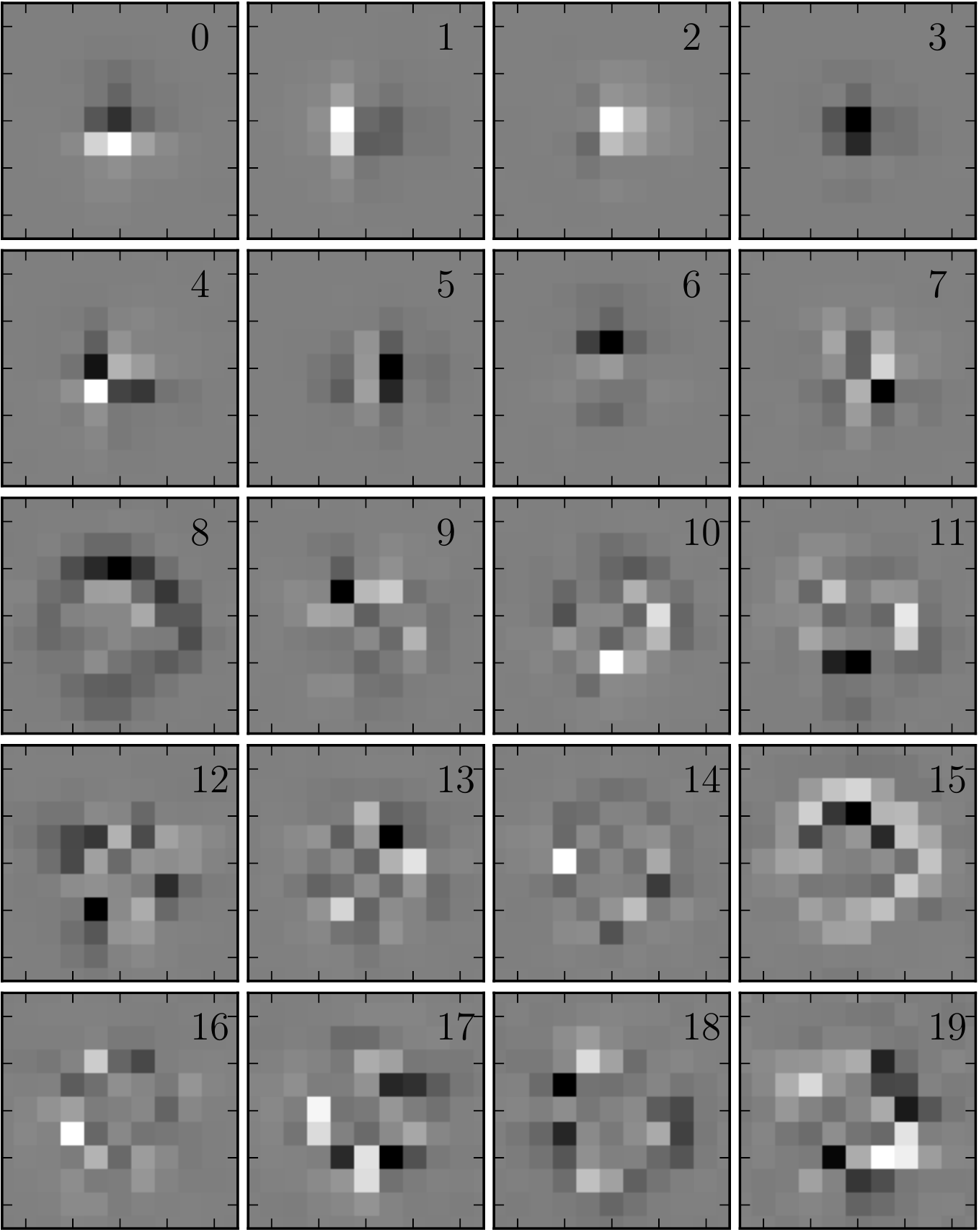}
\caption{\label{fig:eigenstars}Eigen-stars for the \euclid\ PCA decomposition for the first 20 components ($10\times 10$ central pixels). White is positive and black is negative. The first eight components deal with the center of the image while the others describe the wing of the profile.}
\end{figure}

\section{Optimisation of the hyper-parameters}\label{sec:hyperopt}

The performances of machine learning techniques such as neural networks depend on a number of hyper-parameters, for which successful values can be difficult to guess {\it a priori}.
We now describe how we evaluate a grid of possible settings for the hyper-parameters of the classifier, in order to determine optimal configurations. We perform these optimisations only for the \euclid \ and \textit{HST} F606W cases. For the F814W filter, we use the same optimised configuration as for the F606W filter. The hyper-parameters considered here are: the number of retained PCA components $n_\text{PCA}$, the number of hidden layers of the ANN, $n_\text{l}$, and the number of nodes per hidden layer $n_\text{hn}$. 
The capacity of a neural network to learn a task is determined by the values of $n_\text{pca}, n_l$ and $n_\text{hn}$. Large values of the parameters are difficult to train and are prone to over-fitting \citep{Bengio2009}. Small values of the parameters usually result in a somewhat faster training than for large value, but poorer performance, because of under-fitting.

We study the following possible values, whose ranges are determined empirically from preliminary trials:

\begin{eqnarray}
 n_\text{pca}&\in& \{12, 15, 18, 21, 24, 27, 30, 33\}, \label{eq:npca}\\
  n_l&\in& \{2, 3\}, \label{eq:nl}\\
n_\text{hn}&\in& \{5, 8, 11, 14, 17, 20, 23, 26, 29\}.
\end{eqnarray}

For each resulting combination of hyper-parameters, we train 96 ANNs and retain only the $n_c$ best ANNs. The number $n_c\in \{24, 48, 72, 96\}$ is selected to yield the highest $F_1$-score on the validation set. The use of the separate validation set instead of the training set penalises potential over-fitting, although no over-fitting is detected in the present application.
In the context of this paper, we do not systematically explore further hyper-parameters for each setup. In particular, the size of the image stamps on which the PCA is performed is kept constant (40 pixels on-a-side).

\begin{figure}[t!]
\begin{center}
\includegraphics[width=1\linewidth]{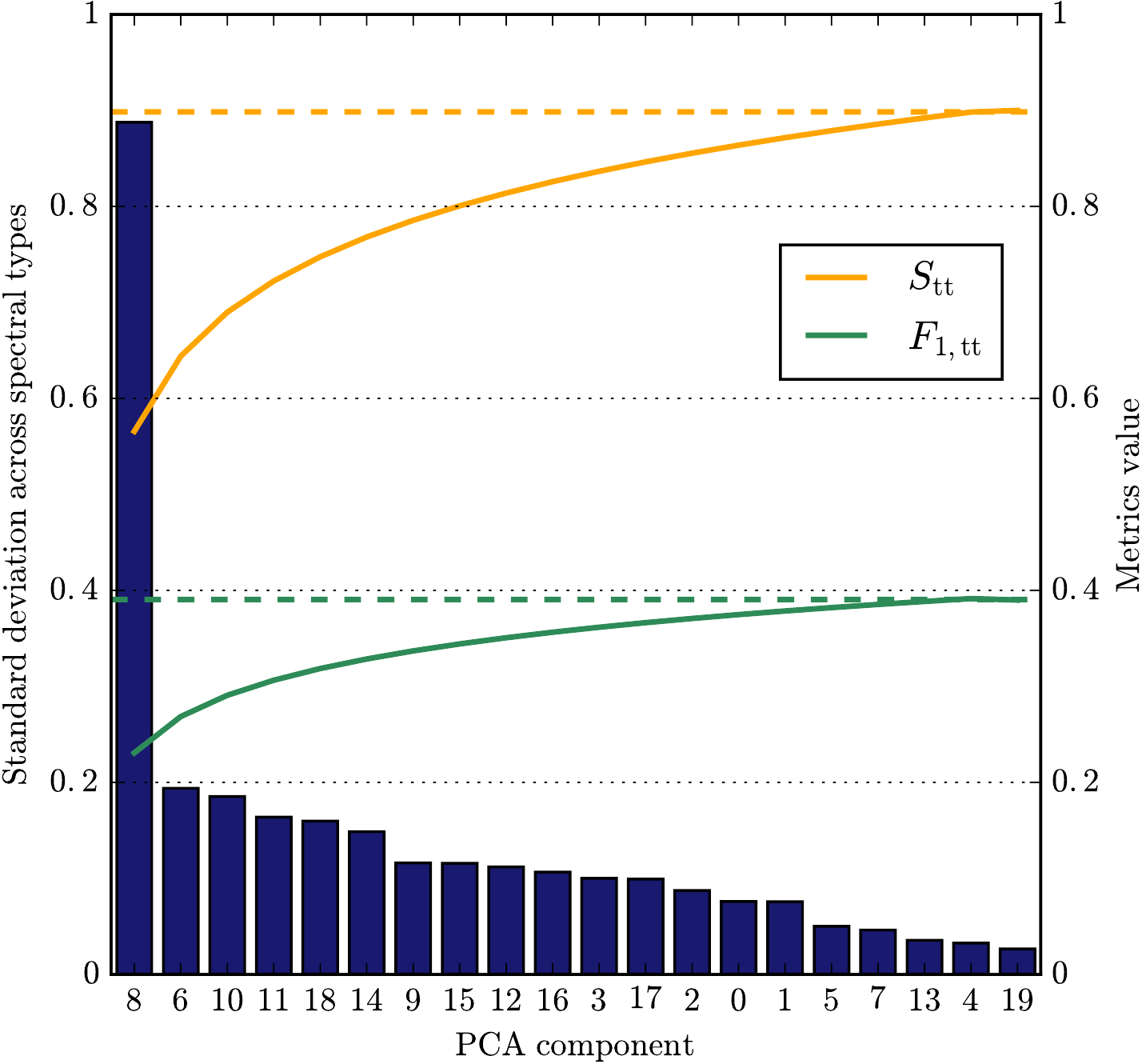}
\caption{\label{fig:ovar_pca} Performance of the classifier with the most significant PCA components and significance of the PCA components. The bar chart shows the standard deviation in the distribution of the first 20 PCA components across the spectral classes in decreasing importance in the context of \euclid\ simulations. The orange and green lines represent the performance metrics $S$ and $F_1$ for classifiers that use only the most ``significant'' PCA components, as defined in the text, and given leftwards in the bar chart. The dashed lines depict the results of the optimisation.}
\end{center}
\end{figure}

Table~\ref{tab:optimisation} presents optimal settings, meaning with the best success rate, $S$, on the validation sets. We stress that the given metrics reflect the performance given the artificial flat distributions of spectral type and S/N, as described in Section \ref{sec:data} and Table~\ref{tab:sets}. 

In Fig.~\ref{fig:configs}, we show the range of performance metrics achieved by the different combinations of hyper-parameters, that is the configuration. The plateaux in the left panels of the figure suggest that the choice of the configuration does not influence much the results and that poor performance of some configurations can easily be identified using the validation set. The same holds true for the number of PCA coefficients used to describe the stellar images. The relatively broad plateaux in the right-hand panels of Fig.~\ref{fig:configs} indicate that this parameter, $n_\text{PCA}$, has only a minor impact on the metrics values. Thanks to this behaviour, a crude optimisation of the hyper-parameters is sufficient.

\subsection{On the significance of the PCA components}

The PCA as described in Section \ref{dimred} is performed on a large ensemble of stars, mixing widely different spatial locations on the detector, different sub-pixel stellar positions, and different spectral classes. We illustrate the first 20 eigen-stars from this PCA, for the \euclid case, in Fig.~\ref{fig:eigenstars}.

Instead of selecting the \emph{first} $n_\text{PCA}$ components as features for the machine learning, one could pick those components that are the most ``significant'' for the purpose of spectral classification. For each PCA component, we quantify this specific significance by evaluating how sensitive the coefficient is to the spectral class when the nuisance parameters (spatial PSF variability, sub-pixel position, noise) are averaged over. To do so, using the same sample of stars on which the PCA was performed, we first compute the median of the eigenvalues for each component and for each true spectral class. For each component, we then compute the standard deviation across these median coefficients from the different spectral classes. The larger this standard deviation, the stronger a PCA component reacts to the morphological differences resulting from the different spectra. This is illustrated in Fig.~\ref{fig:ovar_pca} for \euclid, highlighting the high value of the PCA component number eight in this particular case. We find that selecting the nine most significant coefficients as input features for the network allows us to achieve 90\% of the performance obtained when we use the full 24 coefficients. Even when we use only the single most significant PCA component, the classifier does not lead to catastrophic failures. Adding information from other significant coefficients of course improves the performance.

\begin{figure}[]
\centering
\includegraphics[width=.95\linewidth]{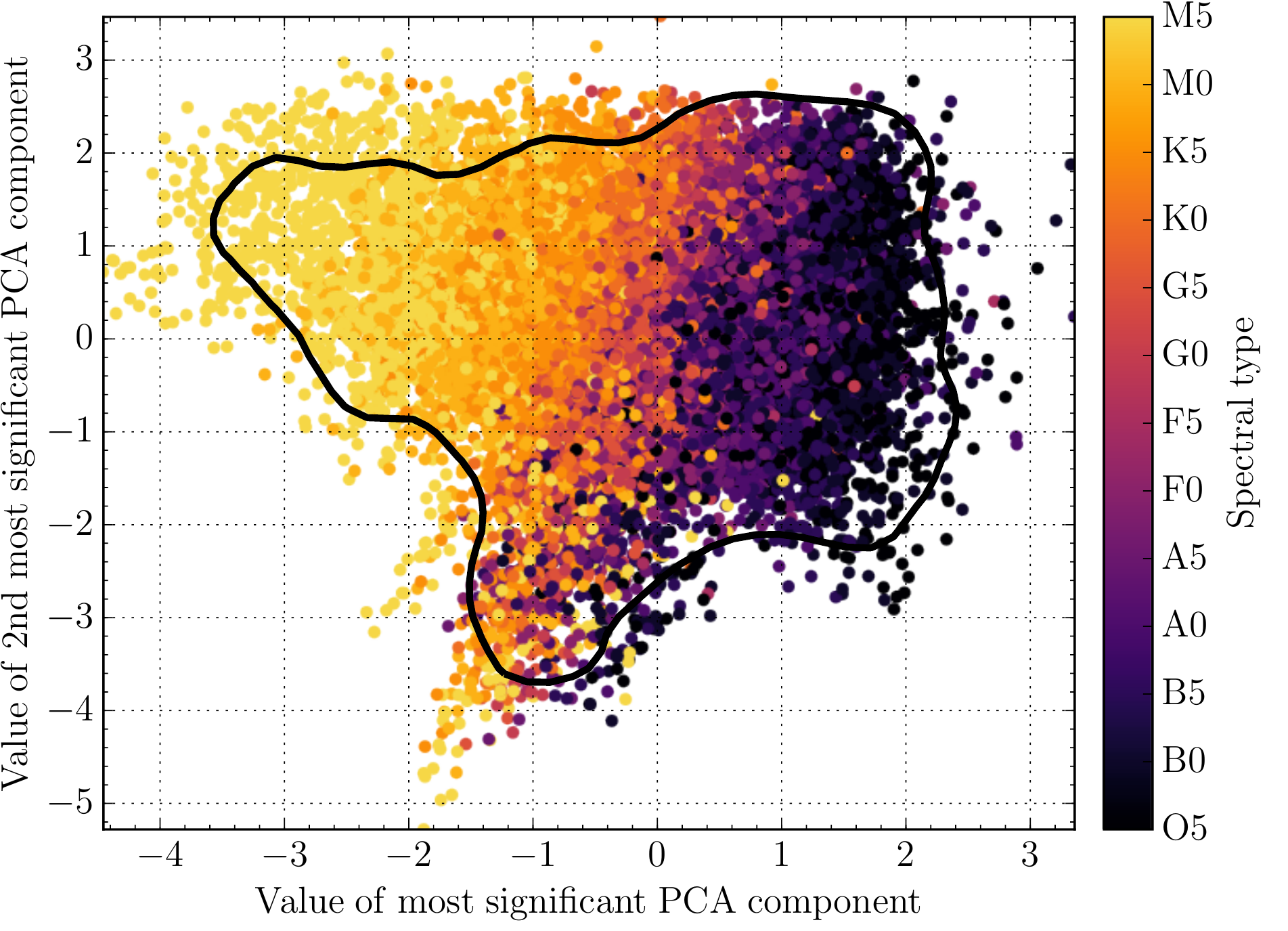}
\caption{\label{fig:pca_decomposition}PCA decomposition of the test set data. The black line depicts the envelope of the PCA decomposition of the training data set. The $x$-$axis$ corresponds to the component with the largest standard deviation (see text) in coefficients across the spectral types. The component with the second largest standard deviation is shown on the $y$-$axis$.}
\end{figure}

Considering Figures \ref{fig:eigenstars} and \ref{fig:ovar_pca}, one can observe that the most significant PCA components usually represent the outer part of the profile, while the first eight coefficients account mainly for the central parts and the centering. Using components eight and six is, for example, an efficient way to measure the slope of the profile. Values of these two components are shown in Fig.~\ref{fig:pca_decomposition}, which illustrates a strong correlation between these coefficients (position of the points in the plot) and the spectral type (color of the datapoints). 

We note, however, that the results presented in the next section, use the optimised value $n_\text{PCA}$ for the number of PCA coefficients to ensure the maximum performance. 

\section{Results} \label{sec:results}

This section presents the performance of the single-band classifier in different conditions, using the optimal configuration as summarised in Table~\ref{tab:optimisation}. 

\begin{figure}[!h]
\begin{center}
\includegraphics[width=1\linewidth]{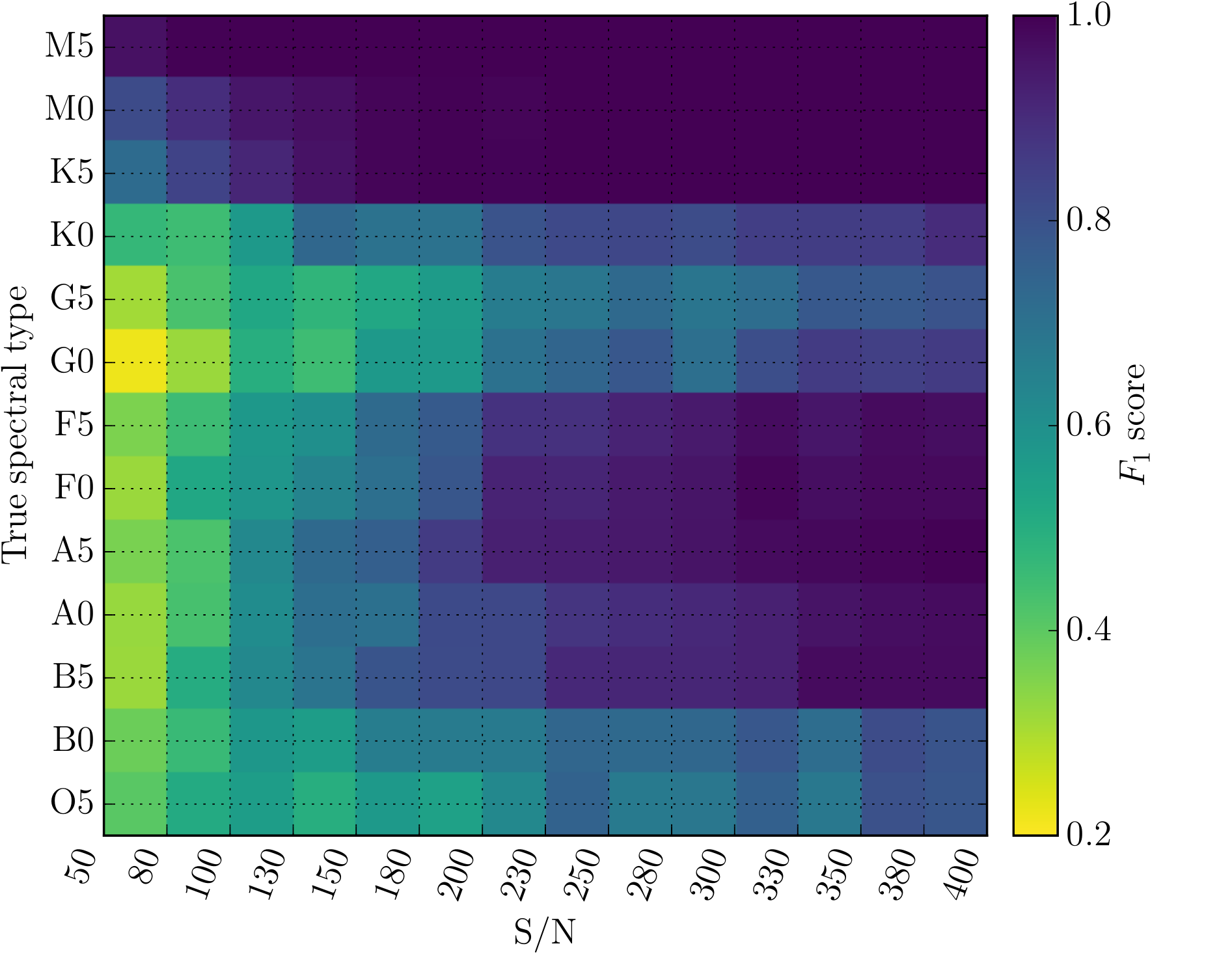}
\caption{\label{fig:c_vs_snr}Classification performance as measured by the $F_1$ score, as a function of the true spectral class and the signal-to-noise ratio, for \euclid.}
\end{center}
\end{figure}

\subsection{Classification results: simple proof-of-concept situation}\label{sec:toy}

\begin{figure*}[tb]
\centering
\includegraphics[width=.75\linewidth]{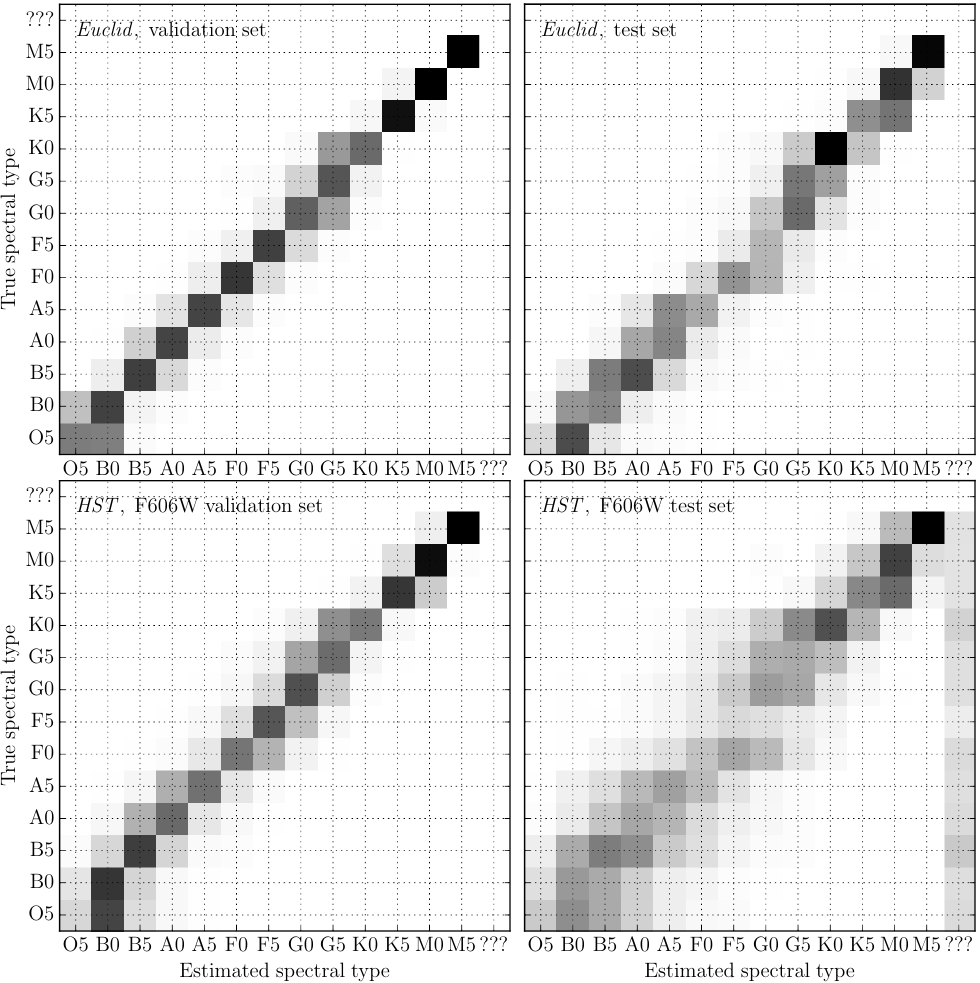} 
\caption{\label{fig:euclid_confusion_mat}Confusion matrices (see Section~\ref{metrics}) for \euclid\ (top) and for the \textit{HST} F606W filter (bottom). The left-hand panels correspond to the training set, while the right-hand panels show results from the more complex test sets. The label ``???'' denotes the ``anomaly'' class (see Section~\ref{sec:spectral_types}).}
\end{figure*}

\begin{figure*}[tb]
\centering
\includegraphics[width=1\linewidth]{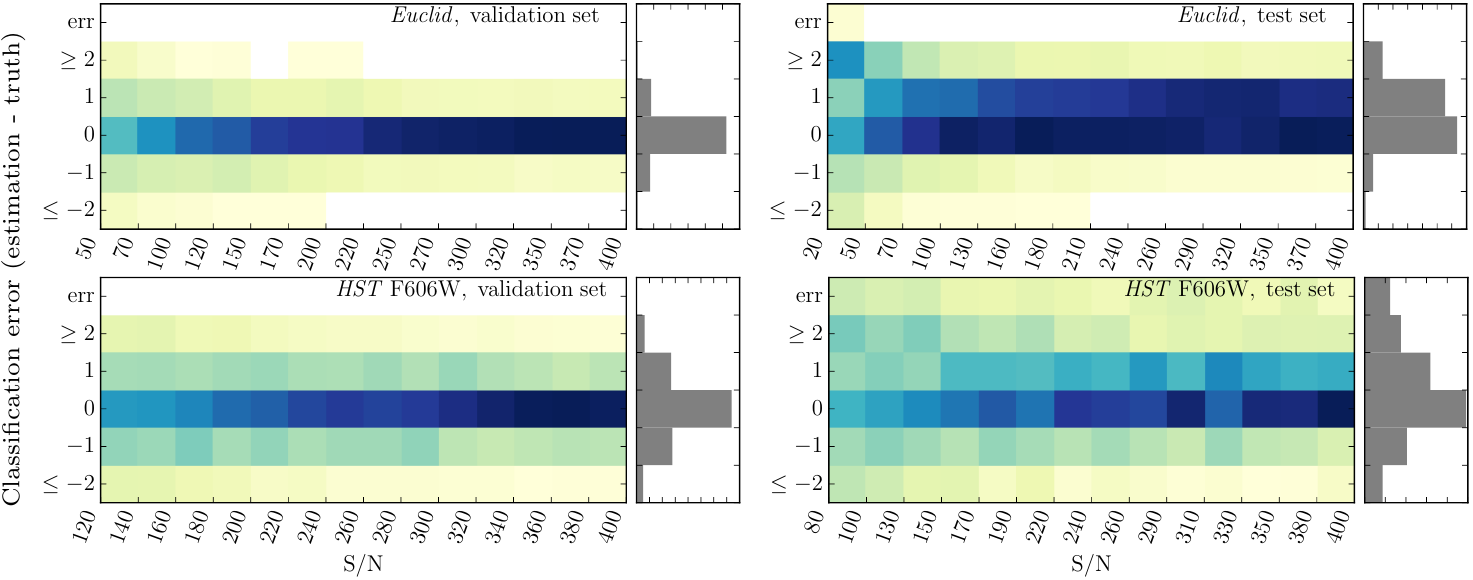} 
\caption{\label{fig:euclid_snr} Classification error as a function of the S/N for \euclid\ (top) and \textit{HST} F606W (bottom). Results from the validation set are shown in the left-hand panels, and results for the test sets are displayed in the right-hand panels. These panels show the same data as Fig.\ref{fig:euclid_confusion_mat}.}
\end{figure*}

We first present results obtained from a simple and well controlled toy model. We use the wavelength-dependent PSF at a single spatial position of the detector to simulate all stellar images, corresponding to a spatially invariant PSF. Furthermore, we use the same S/N ranges and the same stellar spectra for training and testing, and we do not include any extinction effects in the test set.

This simplified situation results in the best possible performance for the problem at hand. For \euclid, and using a uniform distribution of S/N between 50 and 400, we obtain $F_1=0.78$, and a success rate $S=0.99$. The value of $S$ is significantly closer to one than the $F_1$ score as $S$ includes a tolerance of half a spectral class, as compared to the $F_1$ score. This indicates that the vast majority of the classification failures correspond to errors of only half a spectral type.

Figure~\ref{fig:c_vs_snr} shows the $F_1$ score as a function of the true stellar class and of the S/N. The spectral types G0, G5 and K0 present poorer results than their neighbouring spectral classes, reflecting similarity in their spectra. The S/N barely impacts the performance for the reddest objects, but appears more important for bluer objects.

\subsection{Classification results: realistic PSF field}

We now move to the situation of a spatially variable PSF, and we analyse the classification performance on the validation and test sets as described in Section \ref{sec:data}. The analysis of the test sets mimics a data-driven approach, in which the training would be performed on a set of spectroscopically-classified stars with imaging data of higher S/N than for the stars to be classified (the test set).

Figure~\ref{fig:euclid_confusion_mat} shows the confusion matrices for the \euclid\ and the \textit{HST} F606W validation and test sets. We could also consider the \textit{HST} F814W filter but since the spectral slopes of the stars in this filter are similar we expect lower performances. In the following we will only explore the behaviour of our classifier using the bluer F606W filter. For both observational setups, most of the stars are distributed along the diagonal, with a noticeable excess of prediction errors concentrated in the G0 to K0 region, as previously observed for the simpler test described in Section~\ref{sec:toy}.
Also, we observe again that the classification of blue stars (e.g., O5 and B0) is less successful than for the reddest stars.

Figure~\ref{fig:euclid_confusion_mat} shows the degradation of performance between the training and the test phases. The difference can be explained by the inclusion of low S/N images in the test set, as described in Table~\ref{tab:sets}. This degradation, while severe in the F606W filter, still allows for a useful classification of the spectral types, with a typical error of one spectral class. For these F606W simulations, a significant number of stars are classified as anomalies, denoted by the class ``???'' in the figures. In the present case where the test set contains only stars, anomalies are objects that are actually stars but that are classified as not being in the range of objects known by the classifier. The stellar images wrongly classified as anomalies are low S/N objects.

\subsection{Effect of reddening and extinction}

Interstellar dust reddens stellar spectra. In our simulations, we have deliberately included such reddening to the test stars, but not to the training and validation stars. This mimics a situation where the training set is simulated from templates but where the test set has unknown reddening.

Figure \ref{fig:euclid_snr} presents the same results as shown in Fig. \ref{fig:euclid_confusion_mat}, but projected on different axes, namely classification error and S/N. A classification error of $+1$ corresponds to classifying an object as redder than it really is. It becomes apparent that the reddening of the test set results in a bias in the predicted classes, for all S/Ns. However, this can be overcome by including a randomly distributed reddening in the training set. We carried out such an experiment and noticed that the bias disappeared. The performances increased almost at the same level as when the classifiers were run on data sets without any extinction.

\subsection{The effect of contamination by companion objects}

Objects angularly close to the stars degrade the quality of the PCA decomposition and consequently affect the performance of the classifier.

In order to test this, we created an additional test set for the \euclid\ case, containing only double stars (here we do not care if the stars are physically related or not). The contaminating stars are randomly placed in the considered image stamps, with a minimum distance of 1.5 pixels from the main star and they have a random spectral type. The separation of 1.5 pixels corresponds roughly to the FWHM of the \euclid\ PSF. We only simulate contaminants that are fainter than their host stars. Using the new test set but the original training set with single stars, we observe that:

\begin{itemize}
\item The metric $S$ increases with the distance between the main star and its contaminant.
\item Faint contaminants, that is stellar pairs with a flux ratio of larger than two, have little impact on the classification performance.
\item The presence of contaminants increases the fraction of low S/N stars being classified as anomalies.
\end{itemize}

We conclude from this simple study that the general functionality of the single-band classifier is not critically endangered by the astrophysical reality of close companion objects. The presence of companion objects in the training set may, however, severely degrade the performance of the classifier. Depending on the training set selection strategy, the importance of the purity of the training set should be investigated.

\section{Summary and conclusion} \label{sec:summary}

In this paper, we demonstrate the feasibility of inferring the spectral classes of stars from images taken with a space telescope with a single broad-band filter. This single-band classification relies on the wavelength-dependence of the PSF, which leads to small yet significant changes between images of stars with different spectra. We use supervised machine learning to interpret these changes and predict spectral classes. Such a single-band classification can quickly deliver information about stellar types and colours, even in the absence of multi-band photometry or spectroscopic follow-up. Such information may be useful for selecting stars to be used for modelling the wavelength-dependent PSF of, e.g. \euclid. The inner workings of the single-band classifier that we developed for this study can be summarised as follows.

First, we project the stellar images onto a basis obtained from principal component analysis. This reduces the information content of each stellar image to a set of coefficients. Through experimentation, we find that good results are obtained when considering about 25 PCA coefficients from $40 \times 40$ pixels stamps centered on the target stars.
Second, we train committees of feed-forward artificial neural networks to predict the stellar types based on these PCA coefficients. We obtain best results for networks with 2 to 3 hidden layers of 25 to 30 nodes each.

We perform all our analyses with simulated stellar images from several optical setups: {\it HST ACS} using the F606W or F814W filter, and the \euclid\ VIS filter. While we use simple uniform distributions of spectral types and S/Ns, we include the complications of spatially variable PSFs, reddening, and contamination by companion objects. We stress that the purpose of testing these different instrumental and observational conditions is not to compare them, but to demonstrate the general feasibility of the suggested approach. Performing a comparison would require focusing on a particular scientific interest, involving a specific stellar population. 

Our technique is most efficient with broad pass-bands such as the \euclid\ VIS band. However, we show that even the commonly used filters of the \textit{ACS} (F606W and F814W) are broad enough to obtain a reliable stellar classification. This may allow one to use archival {\it HST} data taken in one single band to infer information, for example on the stellar populations of resolved stellar clusters. Still, the goal of the present work is to describe a proof-of-concept classifier. Improvements leading to a full classifier, possibly used for \euclid, may include the following items:

\begin{itemize}
\item The efficiency of the dimensionality reduction could benefit from a better prior centering of the sources, potentially on a finer pixel grid. In the present paper we simulate centering errors as large as half a pixel.
\item The PCA decomposition  could be replaced with a different one, specifically suited to catch the wavelength-dependent features in the PSF, for example wavelets, starlets, shapelets, etc.
\item Any spatial variation of the PSF across the detector could be properly accounted for, and not just marginalised over. This could be achieved by training different classifiers for different locations of the detector, or by using the detector location as input feature to the machine learning.
\item Instead of performing a regression of a continuous parameter whose value encodes the classification, the requested output could be better adapted to the desired use. For example, it might be more meaningful to predict colours instead of spectral types, or to use a softmax regression to obtain probabilities for distinct classes of interest \citep{Nielsen2015}.
\end{itemize}

The results of this method do not depend much on the exact value of the hyper-parameters, which facilitates the optimisation. However, the training strategy is still survey-dependent. For a space telescope, we are fortunate that the PSF can be modelled fairly easily, hence leading to clean and arbitrarily large training sets. 
Another strategy is to train the ANNs on actual stellar images with known spectral types.
This might be a viable strategy for \euclid, given its exceptional PSF stability, the depth of the survey beyond that of {\it Gaia}, and the broad-band VIS filter.

\begin{acknowledgements}
The authors would like to thank R\'emy Joseph for useful discussions as well as Patrick Hudelot, Koryo Okumura and Samuel Ronayette for providing the \euclid\ simulated PSFs.
We are grateful to the anonymous referee for the valuable comments that improved the quality of this work.
This work is supported by the Swiss National Science Foundation. 
MT acknowledges support from a fellowship of the Alexander von Humboldt Foundation.
This research made use of {\tt Astropy} \citep{astropy}, {\tt Matplotlib} \citep{matplotlib} and {\tt Numpy} \citep{numpy}.
\end{acknowledgements}

\bibliographystyle{aa} % style aa.bst
\bibliography{bib} % your references Yourfile.bib

\end{document}